\documentstyle[twocolumn,prl,aps,psfig,floats]{revtex}

\begin{document}
\draft
\preprint{}

\twocolumn[\hsize\textwidth\columnwidth\hsize\csname
@twocolumnfalse\endcsname

\title{Cooling of a lattice granular fluid as an ordering process}
\author{A. Baldassarri$^1$, U. Marini Bettolo Marconi$^{1}$, A. Puglisi$^2$} 
\address{$^1$ INFM Udr Camerino, Univ. di Camerino,
Dip. Matematica e Fisica, Via Madonna delle Carceri I-62032
Camerino, Italy}
\address{$^2$ Universit\'a ``La Sapienza'', P.le A. Moro 2, 00185 Roma, Italy}

\date{\today}
\maketitle
\begin{abstract}

We present a new microscopic model of granular medium to study the
role of dynamical correlations and the onset of spatial order induced
by the inelasticity of the interactions.  In spite of its simplicity,
it features several different aspects of the rich phenomenology
observed in granular materials and allows to make contact with other
topics of statistical mechanics such as diffusion processes, domain
growth, persistence, aging phenomena. Interestingly, while local
observables being controlled by the largest wavelength fluctuations
seem to suggest a purely diffusive behavior, the formation of
spatially extended structures and topological defects, such as
vortices and shocks, reveals a more complex scenario.

\end{abstract}
] \narrowtext

The effort devoted during the last decades to investigate off
equilibrium systems has achieved a series of successes by virtue of a
combination of experiments, numerical simulations, exact theoretical
results and clever phenomenological arguments, but our comprehension
of the area is far from complete. Among these systems granular
materials (GM), i.e.  assemblies of macroscopic particles which
dissipate their energy through inelastic collisions and frictional
forces, have acquired a special rank due to their complex
phenomenology often intriguing and only partly understood
\cite{review}.

The existing theoretical approaches are based either on realistic
descriptions of the grains or on idealized models, which in virtue of
their major simplicity lend themselves to analytic solutions or to
efficient computations. Furthermore such an idealized modeling can
deliberately sacrifice likeness to reality, in order to identify the
physical mechanism responsible for the salient features of the system.
In such a spirit, we shall discuss a minimal model based on the
simplest rule which describes the inelasticity of collisions and local
momentum conservation.

Years ago, S. Ulam showed that an ensemble of elastic particles,
starting from an arbitrary configuration, converged to a Maxwell
equilibrium distribution, postulating a simple redistribution law of
the kinetic energies of randomly selected pairs to simulate the effect
of binary collisions in an elastic gas \cite{Ulam}.  Ben Naim and
Kaprivski (BK), recently, performed a variation over this theme, by
letting the particles endowed with a scalar velocity to dissipate
inelastically a fraction of the relative kinetic energy at each
collision and found that the total kinetic energy decreases
exponentially with time \cite{Bennaim}.  Both models fulfill
Boltzmann's molecular chaos hypothesis, and consequently rule out the
formation of dynamical correlations.  On the other hand computer
simulations have shown the appearance of a shear instability,
i.e. vortices, during the cooling process, before the spontaneous
formation of density clusters.  In an interesting series of papers,
Ernst and collaborators have put forward a mesoscopic theory of these
phenomena, making a connection with phase ordering
kinetics~\cite{Ernst}.
  
In the present work we shall introduce and study a new microscopic
model which preserves the simplicity of the approach {\em \'a la} Ulam,
and displays a complexity similar to that observed in granular
systems.  The focus of our study will be on the statistics of the
velocity field and on its spatial and temporal correlations, stressing
the analogies and the differences with related models aimed to
describe off equilibrium systems.

We introduce our dynamical model by associating a d-dimensional
velocity field $\bf{v_i}$ with each node of a d-dimensional lattice;
at each time step a nearest neighbor pair $(i,j)$ is randomly selected
and the two velocities are updated according to the rule:

\begin{eqnarray}
{\bf v_j'}&=&{\bf v_j}+\Theta({\bf -(v_i-v_j)}\hat \sigma ) \frac{ 1+\alpha}{2}({\bf(v_i-v_j)}\hat \sigma ) \hat \sigma \nonumber\\
{\bf v_i'}&=&{\bf v_i}-\Theta({\bf-(v_i-v_j)}\hat \sigma )\frac{1+\alpha}{2}({\bf (v_i-v_j)}\hat \sigma ) \hat\sigma
\label{rule}
\end{eqnarray}

\noindent
where ${\hat \sigma}$ represents the unit vector pointing from site
${\bf i}$ to ${\bf j}$, $\Theta$ is the Heaviside function which
enforces the kinematic constraint and $\alpha$ the normal restitution
parameter. We shall measure time in the non dimensional number of
collisions per particle.  In each elementary collision (see
Eq.(\ref{rule})) the total linear and angular momentum are conserved,
whereas a fraction $(1-\alpha^2)/4$ of the relative kinetic energy is
dissipated.  The inelasticity of the collisions
has the effect of reducing the quantity $|({\bf v_i-v_j})\hat
\sigma|$, i.e. induce a partial allignement of the velocities.
Hereafter we report the study performed in two dimensions on a
triangular lattice\cite{1d}.

The freely cooling process exhibits striking similarities with the
quench from an initially stable disordered phase to a low temperature
phase in a magnetic system: whereas in a standard quench process
\cite{Bray} one considers the process by which a thermodynamic system,
brought out of equilibrium by a sudden change of an external
constraint, such as temperature or pressure, finds its new equilibrium
state, in a GM one wants to study the relaxation of a fluidized state,
after the external driving force (whose role is to reinject the energy
dissipated by the collisions, keeping the system in a statistically
steady state) is switched-off abruptly at some time $t=0$.
The rotational symmetry of the order parameter ${\bf v_i}$ and the
momentum conserving interaction determine the presence of many
configurations having comparable dissipation rates.  Due to their
competition the system does not relax immediately towards a motionless
state, but displays a phenomenology similar to that observed in a
coarsening process.

%
%
\begin{figure}[tbp]
\centerline{ \psfig{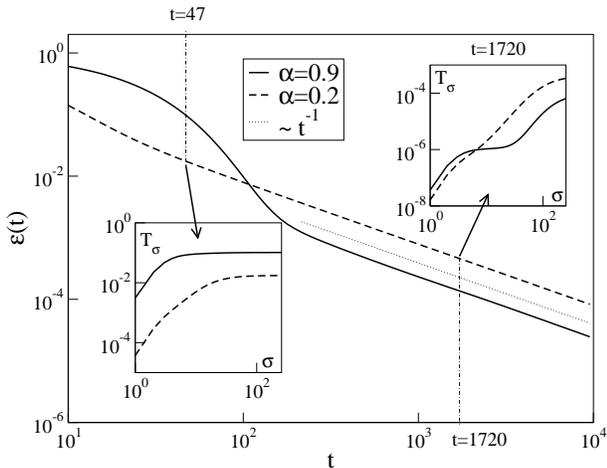} } 
\caption
{Energy decay for $\alpha=0.9$ and $\alpha=0.2$ ($1024^2$ sites). The
dotted line $\sim 1/t$ is a guide to the eye for the asymptotic
energy decay. In the insets we reported the scale dependent
temperature, $T_{\sigma}$, defined in the text, as function of the
coarse graining size $\sigma$ for $t=47$ and $t=1720$. The total
energy per particle and $T_{\sigma}$ remain nearly indistinguishable
in the early incoherent regime, but for {\bf $\sigma<L(t)$} the
thermal energy becomes much smaller than the kinetic energy, a clear
indication of the onset of macroscopic spatial order.}
\label{fig1}
\end{figure}
 
One sees from Fig.\ref{fig1} that during the initial stage, the
total energy per particle $\epsilon (t)=\sum v_i(t)^2/N$ is dissipated
at an exponential rate $\tau^{-1}= (1-\alpha^2)/4$. This can be
deduced from Eq.(\ref{rule}) imposing that each 'spin' fluctuates
independently of the others.  For times larger than $t \sim t_c$, of
the order of $\tau$, the system displays a crossover to a different
regime, where the cooperative effects become dominant and the average
energy per particle decay as $\epsilon(t) \sim t^{-1}$. Such a
behavior agrees with inelastic hard spheres simulations (IHSS) reported in
\cite{Ernst2}.  As shown below, the crossover from one regime to the
other is due to the formation of a macroscopic velocity field. This is
analogous to the formation of magnetic domains in standard quench
processes.  After the formation stage these regions start to compete
to homogenize, causing a conversion of kinetic energy into heat by
viscous heating, i.e. act against the collisional cooling and lead to
a slower decay of the energy~\cite{Ernst}.

Within the early regime the velocity distribution deviates sensibly
from a Maxwell distribution (corresponding to the same average kinetic
energy), but displays fatter tails, a phenomenon which mirrors the
behavior of the BK \cite{Bennaim} model.  The existence of these tails
seems to be due to the lack of spatial correlations, intrinsically
absent at all times in their model, whereas negligible in ours up to
$t_c$. When the energy begins to decay as $t^{-1}$ the velocity
distribution turns Gaussian.  Interestingly, the smaller the
inelasticity, the faster is the energy dissipation, a phenomenon
observed in IHSS \cite{Mcnamara}.


%
%

\begin{figure}[tbp]
\centerline{ \psfig{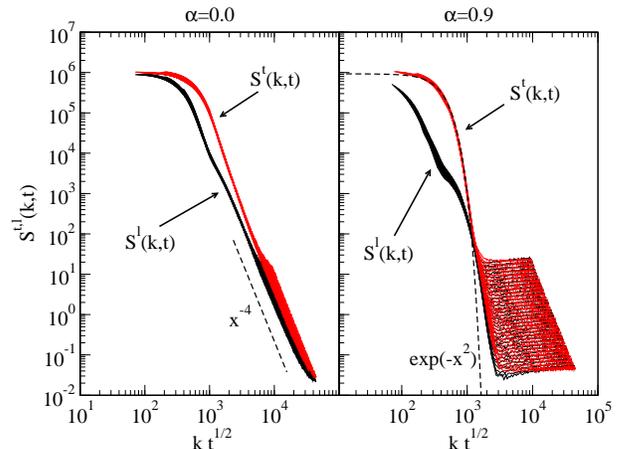} } 
\caption
{ 
Data collapse of the transverse ($S^t$) and longitudinal ($S^l$)
structure functions for $\alpha=0.$ and $\alpha=0.9$ (system size
$1024^2$ sites, times ranging from $t=500$ to $t=10^4$). The
wavenumber $k$ has been multiplied by $\sqrt{t}$. Notice the presence
of the plateaux for the more elastic system. For comparison we have
drawn the laws $x^{-4}$ and $\exp(-x^2)$. }

\label{fig2}
\end{figure}

The most relevant information about the spatial ordering process is
contained in the equal-time structure functions, i.e. the Fourier
transforms of the velocity correlation function:
$$
S^{t,l}(k,t)=\sum_{\hat k} {\bf v^{t,l}}({\bf k},t){\bf v^{t,l}}(-{\bf k},t)
$$
where the superscripts $t,l$ indicate the transverse and longitudinal
components of the field with respect to the wave vector ${\bf k}$ and
the sum $\sum_{\hat k}$ is over a circular shell of radius $k$.  Such
structure factors, if rewritten in terms of the variable $(k t^{1/2})$
display fairly good data collapse, apart from the large $k$ region,
and identify two growing lengths $L^{t,l}(t)$ (see Fig.\ref{fig2}).
Considering the sum rule $\epsilon(t)=\sum_k [S^{t}(k,t)+S^{l}(k,t)]$
we observed that in the early 'exponential' regime the contribution
from the two terms is approximately equal, whereas for times larger
than $t_c$ and $\alpha$ not too small most of the kinetic energy
remains stored in the transverse field, while the longitudinal
component decays faster, with an apparent exponent $t^{-2}$.

The findings concerning the energy decay, the distribution of the
velocity field, and the growth of $L^{t,l}(t)$, lead to the
conclusion, that, if the observation time is longer than the time
between two collisions and if the spatial scale is larger than the
lattice spacing, the system behaves as if its evolution were governed
by a diffusive dynamics~\cite{Ernst}. To be more precise, let us
consider a vector field ${\vec \phi}(x,t)$ which evolves according to
the law $ \partial_t {\vec \phi}=\nu \nabla^2 {\vec \phi} $ starting
from a random uncorrelated initial condition.  The explicit solution
shows that ${\vec \phi}(x,t)$ is asymptotically Gaussian distributed,
with a variance $<\vec \phi(x,t) \vec \phi(x,t)>\propto t^{-d/2}$.
The structure factors $S^{t,l}(k,t)$ assume a scaling form
$S^{t,l}(k,t)=s(kL(t))$ where $L(t)=\sqrt{t}$.  Furthermore, we
compared the two-time autocorrelation $C(t_1,t_2)=\sum_i {\bf
v}_i(t_1) {\bf v}_i(t_2)/N$ with $C_{\phi}(t_1,t_2)=<{\bf
\phi}(x,t_1){\bf \phi}(x,t_2)>$, whose expression reads:
$$
\frac{C_{\phi}(t_1,t_2)}{C_{\phi}(t_1,t_1)}=\frac{2}{(1+\frac{t_1}{t_2} )}
$$

\begin{figure}[tbp]
\centerline{ \psfig{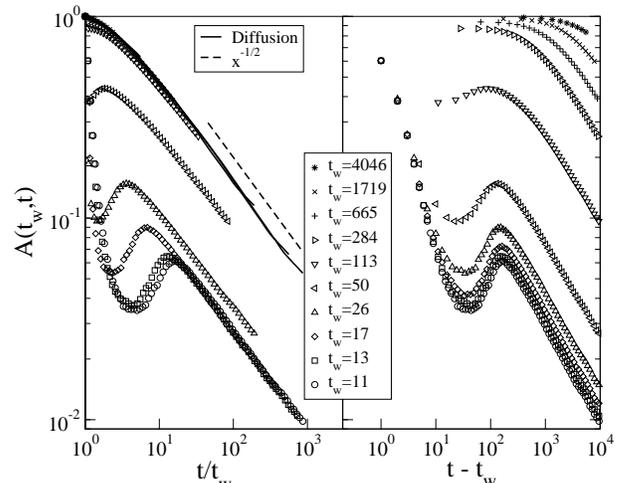} } 
\caption
{ 
Angular autocorrelation function $A(t,t_w)$ for different values of
the waiting time $t_w$ and $\alpha=0.9$ ($1024^2$ sites). The graph on
the left shows the convergence to the $t/t_w$ diffusive scaling
regime, for large $t_w$. For small $t_w$, a local minimum is visible
(for such a quasi elastic dynamics). In the graph on the right the
same data are plotted vs $t-t_w$: note that the small $t_w$ curves
tend to collapse. For higher $t_w$ the position of the local minimum
does not move sensibly, but its value grows and goes to $1$ for large
$t_w$.  }
\label{fig3}
\end{figure}

During a short time transient, the autocorrelation function of our
model differs from $C_{\phi}$, since it depends on $t_1-t_2$, i.e. it
is time translational invariant ({\em TTI}).  Later, $C(t_1,t_2)$
reaches the ``aging'' regime and depends only on the ratio
$x=t_1/t_2$.  Something similar occurs in a coarsening process, where
the autocorrelation of the local magnetization $a(t_w,t_w+\tau)$
reaches, for large $\tau$ (but $\tau << t_w$), a constant value
$m_{eq}^2(T)$, that is the square of the equilibrium
magnetization. Obviously, for $T\to 0$, $m_{eq}^2\to 1$ and the {\em
TTI} transient regime disappears.  The short time transient in our
model is analogous to  such a {\em TTI} regime, with
the difference that the cooling process imposes a decreasing
temperature $T(t_w)\to 0$, that progressively erodes the {\em TTI}
regime. The same dependence on the {\em TTI} manifests itself in the
angular auto-correlation:
$$
A(t,t_w)=\frac{1}{N}\sum_i \cos(\theta_i(t+t_w)-\theta_i(t_w)).
$$
Again, for large waiting times $t_w$ this function assume the
diffusive $t/t_w$ scaling form, but for a small fixed $t_w$, displays
a minimum and a small peak before decreasing at larger $t$ (see
Fig.\ref{fig3}). The non-monotonic behavior of $A(t,t_w)$ suggests
that the initial direction of the velocity induces a change in the
velocities of the surrounding particles, which in turn generates,
through a sequence of correlated collisions, a kind of retarded field
oriented as the initial velocity.
As $t_w$ increases the maximum is less and less pronounced.

%
%

In spite of these first results, that seem to give support to the idea
that the model dynamics is purely diffusive~\cite{persistence}, the
model is more complex.  The main evidence stems from the following
facts:

i) the structure functions do not have the typical Gaussian tails of a
diffusive system, due to the non linearity represented by the
kinematic factor in Eq.(\ref{rule}) and the shapes of $S^{t,l}(k,t)$
display three different regions: a long-wavelength region which is
diffusive in character; an intermediate region where the structure
functions decay as $k^{-\beta}$ with $\beta \sim 4$; a plateau region
where $S^{t,l}$ decay in time with a power law $t^{-2}$ but remain
nearly constant with respect to $k$ (for $r>0$);

ii) the Fourier modes interact and an initial shear state, obtained
assuming the initial configuration to be a plane wave, decays into
shorter wavelength modes by a mechanism of period doubling; that is to
say, contrary to the diffusion, plane waves are not eigenmodes.

The existence of the quasi-elastic plateaux is the fingerprint of
localized fluctuations which, for small inelasticity, propagate and
are damped less than exponentially.  A small $\alpha$ determines a
rapid locking of the velocities of neighboring elements to a common
value, while in the case of $\alpha\to 1$, short range small amplitude
disorder persists within the domains, breaking simple scaling of
$S^{t,l}$ for large $k$ and having the effect of a self induced noise.

One can characterize such internal noise by means of an average local
granular temperature $T_{\sigma}$, i.e. a measure of the variance of
${\bf v_i}$ with respect to the local average of ${\bf v}$ within a
region of linear size $\sigma$.  Obviously, since when $\sigma \to
\infty$ the local average tends to the global (zero) momentum, then
$T_{\sigma} \to \epsilon$, as shown in the insets of Fig.\ref{fig1}.
For $\sigma < L(t)$, instead, $T_{\sigma} < \epsilon $. For quasi
elastic systems $T_\sigma$ exhibits a plateau for $1\ll \sigma\ll
L(t)$ that identify the strength of the internal noise.  The local
temperature ceases to be well defined for smaller $\alpha$ due to the
absence of scale separation between microscopic and macroscopic
fluctuations in the strongly inelastic regime \cite{Mcnamara}.  The
existence of a $L^{-2}(t)k^{-4}$ region in the structure functions is
consistent with Porod's law~\cite{Bray} and is the signature of the
presence of vortices, a salient feature of the cooling process.
Vortices form spontaneously and represent the boundaries between
regions which selected different orientations of the velocities during
the quench and are an unavoidable consequence of the conservation laws
which forbid the formation of a single domain.  With the random
initial conditions adopted, vortices are born at the smallest scales
and subsequently grow in size by pair annihilation, conserving the
total charge.  By locating the vortex cores, we measured the vortex
density $\rho_v(t)$, which represents an independent measure of the
domain growth, and in fact it decays asymptotically ($t \gg t_c$) as
an inverse power of time, i.e. $L_v(t)=\rho_v^{-1/2} \propto t^{1/2}$.
The vortex distribution turns out to be not uniform for $\alpha$ not
too small. Its inhomogeneity is characterized by the correlation
dimension $d_2$ : $H(R)=N_v^{-2} \sum_{i<j}\Theta(R-({\bf r_i-r_j }))
\sim R^{d_2}$ where the $r_i$ are the core locations.  For $\alpha \to
1$ the vortices are clusterized ($d_2<2$) i.e. do not fill
homogeneously the space, whereas at smaller $\alpha$ their
distribution is homogeneous ($d_2 \to 2$).

%
%

\begin{figure}[tbp]
\centerline{ \psfig{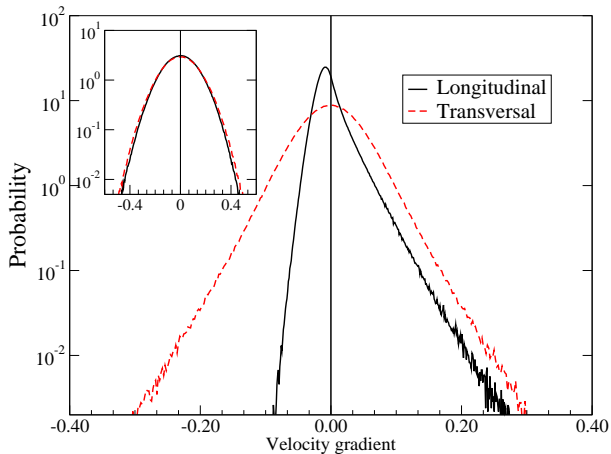} } 
\caption
{ 
Probability densities of the longitudinal and transverse velocity
increments. The main figure shows the p.d.f. of the velocity gradients
($R=1$). The inset shows the Gaussian shape measured for $R=40$
(larger than $L(t)$ for this simulation: $\alpha=0.2$, $t=620$, system
size $2048^2$).  }
\label{fig4}
\end{figure}

Vortices are not the only topological defects of the velocity
fields. In fact we observe shocks, similarly to recent experiments in
rapid granular flows\cite{Swinney}.  Shocks have a major influence on
the statistics of velocity field, i.e. on the probability
distributions of the velocity increments.  The probability density
function (p.d.f.) of the longitudinal increment
\[
\Delta_l({\bf R}) = \frac 1N \sum_{\bf i} ({\bf v_{i+ R}} - {\bf v_i})\cdot \frac {{\bf R}}R
\]
is shown in Fig.\ref{fig4} for $R=1$ (longitudinal velocity gradient)
in the main frame, and for $R=40>L(t)$ in the inset.  For small $R\ll
L(t)$ the longitudinal increment p.d.f. is skewed with an important
positive tail, whereas for $R\gg L(t)$ it turns Gaussian.  The
distribution of transverse increments $({\bf v_{i+ R}} - {\bf
v_i})\times {\bf \hat R}$, instead, is always symmetric, but non
Gaussian distributed for small $R$. A similar situation exists in
fully developed turbulence \cite{multiscal}.

To conclude, our model provides a link between the microscopic rules
of granular dynamics and its hydrodynamical description.  It allows to
follow the cooling of a granular material and the build up of velocity
correlations, by means of efficient numerical measures of structure
factors, two-time correlations and topological defects.  The data
analysis reveals the presence of vortices, shocks and internal noise
and suggests the existence of a scale separation only in the case of
quasi elastic systems, which is instead suppressed for large
inelasticities.

Even independently from the problem of granular flows, the model
represents a simple but unusual phase ordering system.  In fact,
despite the apparently purely diffusive aspects shown by one-point
quantities, it displays anomalous statistics of spatial properties for
the order parameter field as witnessed by the velocity gradient
p.d.f. and by the structure functions.

{\bf Acknowledgments.} We thank E.~Caglioti, M.~Cencini,
L.F.~Rull, A.~Vulpiani and S.~Zapperi for several useful discussions.


\begin{references}

\bibitem{review} H.M.~Jaeger, S.R.~Nagel and R.P.~Behringer,
Rev.Mod.Phys. {\bf 68}, 1259 (1996); T.~P\"oschel, S.~Luding (Eds.)
{\em Granular Gases}, Springer, Berlin (2001).

\bibitem{Ulam} S. Ulam, Adv.Appl.Math. {\bf I}, 7-21 (1980).

\bibitem{Bennaim} E. Ben-Naim, P.L. Krapivsky, Phys.Rev. E, {\bf
61} R5 (2000).


\bibitem{Sela} N. Sela and I. Goldhirsh, Phys.Fluids {\bf 7}, 507
(1995).

\bibitem{Bennaim-1d} E. Ben-Naim, S.Y. Chen, G.D. Doolen, and
S. Redner, {\em Phys.Rev.Lett.} {\bf 83} 4069-4072 (1999).

\bibitem{Ernst}
T.P.C. van Noje, M.E. Ernst, R. Brito and J.A.G. Orza,
Phys.Rev.Lett. {\bf 79}, 411 (1997). 

\bibitem{Ernst2} J.A.G. Orza, R. Brito, M.H. Ernst, {\tt cond-mat/002383}.

\bibitem{Mcnamara} S.McNamara, Phys.Fluids A 5, 3056 (1993).

\bibitem{1d} The 1-d version of our
model compares well with models of hard rods (see
\protect{\cite{Bennaim-1d}}), since in both cases each particle collides only with its
neighbors (A.Baldassarri, U. Marini Bettolo
and A.Puglisi, in preparation).

\bibitem{Bray} A.J. Bray, Adv.Phys. {\bf 43} , 357 (1994).

\bibitem{persistence} Even the persistence exponent $\theta$, defined
through the decay $N_s \sim t^{-\theta}$ of the number of sites where a velocity component
never changed its sign up to time $t$, agrees
with the analogous exponent of the diffusion equation (we measured $\theta=0.18$).
    
\bibitem{multiscal} R. Benzi, L. Biferale, G. Paladin, A. Vulpiani,
and M. Vergassola, Phys.Rev.Lett. {\bf 67}, 2299 (1991).

\bibitem{Swinney} E. Rericha et al.  {\tt cond-mat/0104474}.

\bibitem{Mcnam2} S. Aumaitre, S. Fauve, S. McNamara and P. Poggi,
Eur.Phys.J. B {\bf 19}, 449 (2001).
\end{references}
\end{document}